\documentclass[prb,twocolumn,showpacs,showkeys,superscriptaddress,nosortaddress,preprintnumbers,amsmath,amssymb]{revtex4}
\usepackage{graphicx}% Include  files
\usepackage{dcolumn}% Align table columns on decimal point
\usepackage{bm}% bold math

\usepackage[english]{babel}

\begin{document}

\preprint{PROSHIN}

\title{Hierarchy of critical temperatures
in four-layered \\
ferromagnet/superconductor nanostructures and control devices}% Force line breaks with \\

%\title{Four-layered ferromagnet/superconductor nanostructures:\\
%hierarchy of critical temperatures and control devices}

%\title{Four layered ferromagnet - superconductor nanostructures: \\
%a hierarchy of critical temperatures}% Force line breaks with \\

\author{Yurii N. Proshin}
%\affiliation{Kazan State University, Kazan, Russia }
 \email{Yurii.Proshin@ksu.ru}
\affiliation{Kazan State University, Kazan, Russia }
\affiliation{Max-Planck-Institute
for the Physics of Complex Systems, Dresden, Germany}%Lines break automatically or can be forced with \\
%\altaffiliation[Present address: ]{Max-Planck-Institute
%for the Physics of Complex Systems, Dresden, Germany}%Lines break automatically or can be forced with \\

  \author{Alexei Zimin}%
\affiliation{Kazan State University, Kazan, Russia }
 \author{Nail G. Fazleev}%
\affiliation{Kazan State University, Kazan, Russia }
\affiliation{University of Texas at Arlington, Arlington, USA}
 \author{Mansur G. Khusainov}%
\affiliation{Kazan State University, Kazan, Russia }
\affiliation{Max-Planck-Institute
for the Physics of Complex Systems, Dresden, Germany}%Lines break automatically or can be forced with \\
\affiliation{``Vostok'' branch, Kazan State Technical University,
Chistopol', Russia }

%\date{\today}% It is always \today, today,
\date{Received 11 February 2005; published 11 May 2006}             %  but any date may be explicitly specified

\begin{abstract}
The four-layered F/S/F$'$/S$'$ nanostructure consisting of rather
dirty superconducting (S) and ferromagnetic (F) metals is studied
within the theory of the proximity effect taking detailed account
of the boundary conditions. The F/S structures with four F and S
layers are shown to have considerably richer physics than the
F/S/F trilayer (due to the interplay between the 0 and $\pi$ phase
superconductivity and the 0 and $\pi$ phase magnetism) and even
the F/S superlattices. The extra $\pi $ phase superconducting
states obtained for the four-layered F/S/F$^{\prime }$/S$^{\prime
}$ system are found to be different from the known
``superlattice'' states. The dependence of the critical
temperatures versus the F layers thicknesses is investigated. An
optimal set of parameters is determined, for which the difference
between the critical temperatures for different states becomes
significant, and the corresponding phase diagrams are plotted. It
is proven that this system can have different critical
temperatures for different S and S$'$ layers. A conceptual scheme
of a control device with superconducting and magnetic recording
channels that can be controlled separately using a weak external
magnetic field is proposed on the basis of the F/S/F$'$/S$'$
nanostructure. The devices with four, five, six, and seven different states
are explored.
\end{abstract}

% PACS, the Physics and Astronomy
                             % Classification Scheme.
%Use showkeys class option if keyword
                              %display desired

\maketitle

%\reversemarginpar

\section {Introduction}

Superconductivity and ferromagnetism are antagonistic ordering
phenomena and their coexistence in homogeneous materials requires
special conditions that are hard to realize. One of possible
explanation of the superconductivity suppression by ferromagnetic
ordering in transition metals was given by~\citet{Ginzburg}, who
noted that the magnetic induction exceeds the critical field. This
antagonism is also clear from the viewpoint of the microscopic
theory: the attraction between electrons creates Cooper pairs in a
singlet state, whereas the exchange interaction producing
ferromagnetism tends to arrange electronic spins parallel to each
other. Therefore, when the Zeeman energy of the electrons of a
Cooper pair in the exchange field $I $ exceeds the coupling
energy, the measure of which is the superconducting gap $\Delta $,
the superconducting state is destroyed. In contrast to critical
field $H_{\text{c}} $ acting on orbital states of the electrons of
a pair, the exchange field acts on electronic spins (spin degrees
of freedom), therefore the destruction of superconductivity due to
this field is called the paramagnetic
effect~\cite{Chandra,Clogston}.

However, the above mentioned coexistence of superconducting and
ferromagnetic order parameters is easily achievable in fabricated
or natural F/S heterostructures consisting of alternating
ferromagnetic metal (F) and superconducting (S) layers. In this
case \emph{superconducting} and \emph{ferromagnetic} electronic
systems are \emph{spatially} separated. Due to the proximity
effect, the superconducting order parameter can be induced in the
F layer; on the other hand, the neighboring pair of the F layers
can interact with each other via the S layer. Such systems exhibit
rich physics, which can be controlled by varying the thicknesses
of the F and S layers, or by placing the F/S structure in an
external magnetic field.

The modern technologies of production of the layered structures,
such as molecular-beam epitaxy, allow to deposit layers of atomic
thickness and to study the properties of such heterogeneous F/S
systems as a function of the ferromagnetic ($d_{\text{f}} $) or
superconducting ($d_{\text{s}}$) layer thickness. Numerous
experiments on the F/S structures (contacts, trilayers, and
superlattices) have revealed nontrivial dependences of the
superconducting transition temperature $T_c$ on the thickness of
the ferromagnetic layer (see the
reviews~\cite{OurUFN02,Ryasanov04,Buzd_RevModPhys05} and
references therein).

The boundary value problem for the pair amplitude (the Cooper pair
wave function) in a dirty superconductor for the F/S superlattice
was formulated in pioneering works by~\citet{Rad91,Buzd92}. The
critical temperature $T _ {\text{c}} $ that was also calculated as
the $d _ {\text{f}} $ function in Refs.~\onlinecite{Rad91,Buzd92}
exhibited both monotonic and nonmonotonic dependences.
Oscillations of $T _ {\text{c}} (d _ {\text{f}}) $ were related to
periodical switching of the ground superconducting state between
the 0 and $ \pi $ phases, so that the system chooses the state
with higher transition temperature $T_{\text{c}}$. In the $\pi$
phase state the superconducting order parameter $\Delta$ in the
neighboring S layers of the F/S superlattice have the opposite
sign, contrary to the 0 phase state in which $\Delta$ has same
sign for all S layers. The experimental evidence of the $\pi$
superconducting state in the F/S systems has been discussed in the
review~\cite{Ryasanov04}. The concept of a $\pi$ junction was
proposed by~\citet{Bulaev77}.
%
%\marginpar{Comm_1 without ``really''}

However, the boundary conditions used in
Refs.~\onlinecite{Rad91,Buzd92} are correct only in the limit of
high transparency of the F/S interface. In subsequent
studies~\cite{OurPRB,OurJETP_L,OurJETP,Aarts97} the boundary
conditions have been derived from the microscopic theory, and they
are valid for arbitrary transparency of the F/S interface. The
solution of the boundary value
problem~\cite{OurPRB,OurJETP_L,OurJETP,Aarts97,LTag98,FominovJETPL01,FominovPRB02,BagretsPRB03}
has revealed an additional mechanism of nonmonotonic dependence of
$T_{\text{c}}$ due to modulation of the pair amplitude flux from
the S layer to the F layer. This modulation is caused by the
change of the FM layer thickness $d _ {\text{f}}$. Moreover, it
has also resulted in a prediction of different types of behavior
$T_{\text{c}}(d_{\text{f}})$ such as
reentrant~\cite{OurPRB,OurJETP_L,OurJETP,OurJETPL00} and
periodically reentrant
superconductivity~\cite{OurPRB,OurJETP_L,OurJETP}. Note that both
the oscillations and the reentrant behavior of $T _ {\text{c}} (d
_ {\text{f}}) $ can appear not only in the F/S superlattice but
also in simple F/S bilayer and F/S/F trilayer systems in which the
$ \pi $ phase superconductivity is impossible in principle! The
reentrant character of superconductivity that we have predicted
has been recently observed experimentally in the Fe/V/Fe
trilayer~\cite{Garif02}.

Now it may be considered as proven~\cite{OurUFN02} that
superconductivity in the layered F/S systems is a combination of
the BCS pairing with a zero total momentum of the pairs in the S
layers and the pairing due to the Larkin-Ovchinnikov-Fulde-Ferrell
(LOFF) mechanism~\cite{LO,FF} with a nonzero three-dimensional
(3D) momentum of the pairs ${\bm k}$ in the F layer. The LOFF
pairs momentum ${k \simeq 2I/v_f}$ is determined by the Fermi
surface splitting caused by the internal exchange field $I$ (where
$v_f$ is the Fermi velocity in the F layers). Usually it is
assumed~\cite{Rad91,Buzd92,Demler,OurJETP,OurJETP_L,OurPRB,Aarts97,FominovJETPL01,FominovPRB02,BagretsPRB03,LTag98}
that the momentum of  the LOFF pairs is directed across the F/S
interface (the so-called one-dimensional (1D) case). In our recent
papers~\cite{OurJETPL00,OurUFN02,OurUFN03} we took into account
the spatial variations of the pair amplitude not only across the
F/S nanostructure but also along the F/S boundary (the 3D case).
In the general case, this leads to the increase of the critical
temperature $T_{\text{c}}$ and to the smoothing of the $T _
{\text{c}} (d _ {\text{f}}) $ oscillations, in comparison with the
1D version of the theory, due to the 3D-1D-3D phase transitions.
The appearance of the 3D-1D-3D phase transition cascade is
associated with the umklapp processes at which the LOFF pairs
momentum  ${\bm k}$ is exactly conserved up to a minimal
reciprocal lattice vector ${\bm g}$ of the 2D surface LOFF states.
Therefore ${\bm k}$ is actually a quasimomentum and this fact is
reflected in the revised F/S boundary
conditions~\cite{OurUFN02,OurUFN03}. The use of the latter in turn
can result in a $T _ {\text{c}} (d _ {\text{f}}) $ dependence with
one local minimum, which is a typical experimental nonmonotonic
behavior~\cite{OurUFN02}.

Of special interest is the study of the multilayered F/S
structures, in which various types of magnetic order can arise in
the F layers due to their indirect interaction via the S layers.
Recently the theory of the proximity effect has been developed for
the F/S structures taking into account the inverse influence of
superconductivity on magnetism of the F layers and on mutual
orientation of their magnetizations. This aspect of the proximity
effect has been studied for the F/S/F trilayer
``spin-switch"~\cite{Buz_Ved,LTag99} and exploring the possibility
of the cryptoferromagnetic state in the
F/S~bilayer~\cite{Berg00,Berg04PhRvB..69q4504B}. The long-range
proximity effect due to triplet superconductivity that arises in
the case of non-collinear alignment of magnetizations in the F
layers has been studied for the F/S/F trilayer system  in
Refs.~\onlinecite{Berg01PhRvL..86.4096B,Berg03PhRvB..68f4513B,Volkov03PhRvL..90k7006V,FominovJETPL03}.

An interplay between the 0 and $\pi $ phase types of
superconductivity in the S layers should be included in the
above-mentioned magnetic mutual accommodation in the F/S
\textit{superlattices}. This added competition leads to two
layered antiferromagnetic superconducting ({\sf AFMS})
states~\cite{OurUFN02,OurJETPL01,OurPRB01}. In the {\sf AFMS}
state the phases of the magnetic order parameter in the
neighboring F layers are shifted by $\pi $, i.e. the exchange
fields $I$ have opposite signs in the neighboring F layers. This
state with antiparallel alignment of the corresponding
magnetizations may be considered as a manifestation of the $\pi$
phase magnetism. Similar to the F/S/F
\textit{trilayer}~\cite{Buz_Ved,LTag99}, in the case of the F/S
superlattice the \textsf{AFM} ordering of the magnetizations of
all F layers  leads to the significant \textit{reduce} of the
pair-breaking effect of the exchange field $I$ for the S layers,
and to the \textit{raise} of the critical temperature of the
layered system. This theoretical prediction has been
experimentally confirmed for the Gd/La
superlattices~\cite{Goff02}. \citet{Goff02} have observed that the
superlattices with prepared antiferromagnetic ordering of the
magnetizations in the adjacent Gd layers undergo the  transition
into a superconducting state at considerably higher temperatures
in comparison with the superlattices with ferromagnetic ordering
of the Gd layers. This mutual accommodation between the
superconducting and magnetic order parameters reflects a quantum
coupling between the boundaries. The competition between the 0 and
$\pi$ phase superconductivity and the 0 and $\pi$ phase magnetism
leads to a change in the classification of the F/S superlattice
state~\cite{OurJETPL01,OurPRB01,OurUFN02}.

The F/S nanostructures possess two data-recording channels: the
superconducting one determined by conducting properties of the S
layers, and the magnetic one determined by ordering of the F layer
magnetizations. The F/S/F \textit{trilayer} devices, proposed in
Refs.~\onlinecite{Buz_Ved,LTag99,Oh97}, operate through
transitions between the superconducting (\textsf{S}) and normal
(\textsf{N}) states that are induced by changes of the mutual
ordering of the magnetizations of the adjacent FM layers. These
changes are controlled by an external magnetic field $H$. The data
stored in the superconducting and magnetic channels of this
\textit{switch} device change \textit{simultaneously}, and the
magnetic order completely determines the ``superconducting
information". The scheme of a complex device on the basis of the
F/S \textit{superlattices}, in which the superconducting and
magnetic data-recording channels can be controlled separately, has
been proposed in Refs.~\onlinecite{OurJETPL01,OurPRB01}.

In Section II we briefly discuss the earlier
proposed control devices (``spin switches'') based
on the F/S nanostructures. In Section III we explore
the four-layered F/S/F$'$/S$'$ system (see
Fig.~\ref{fig1}) assuming the competition between
the 0 and $\pi $ phase magnetism and the 0 and $\pi
$ phase superconductivity takes place. We solve the
Usadel equations for this structure taking into
account the boundary conditions. In Section IV we
construct the phase diagrams with an optimal set of
parameters. In Section V we propose a scheme of a
control device based on the studied F/S/F$'$/S$'$
system and discuss its few possible operating
regimes.

\section {Spin switches for current on the  basis of F/S heterostuctures}

A conceptual scheme of a spin switch device for
current based on a F/S/F \textit{trilayer} was
proposed by~\citet{Buz_Ved} and \citet{LTag99} for
the case of the ``Cooper limit'' when the
thicknesses $d _ {\text{s}} $ and $d _ {\text{f}} $
of the S and F layers are much less than the
corresponding coherence lengths $ \xi _ {\text{s}} $
and $ \xi_ {\text{f}} $, respectively. It has been
theoretically shown, that the ``antiferromagnetic''
({\sf AFM}) configuration of such a three-layered
system with an antiparallel arrangement of the
magnetizations of the F layers has a higher
transition temperature $T _ {\text{c}} $ in
comparison with the one for the ``ferromagnetic''
({\sf FM}) configuration. In other words the {\sf
AFM} configuration  is energetically more favorable
and the {\sf AFMS} state is the ground state of this
system at $T < T _ {\text{c}}$ in the absence of an
external magnetic field. The nature of this behavior
of $T _ {\text{c}}$ is related to a reduction of the
pair-breaking action of the exchange field of the F
layers in the {\sf AFM} configuration on
superconducting pairs, i.e. to a partial
compensation of the paramagnetic effect.

Applying small magnetic field $H$ higher than the coercivity
$H_{\text{coer}}$ of the F layer one can change the \textsf{AFM}
orientation of the magnetizations to the \textsf{FM} one. Under
certain conditions the trilayer system can undergo a transition
from a superconducting ({\sf AFMS}) state to a normal
(\textsf{FMN}) one, i.e. from a state with zero resistance to a
resistive one. As the magnetic field is turned off, the {\sf AFM}
orientation of the magnetizations (the $\pi $ phase magnetic
state) and the superconducting properties of the system are
restored. Note that in a certain sense the F/S/F switch operates
in the same manner as an usual isolated superconductor which turns
into the normal state if the applied field $H$ exceeds the
critical field $H_{\text{c}}$. Moreover, since the value of the
critical field $H _ {\text{c}} $ is determined by the difference
between the critical temperature and the temperature of a sample
$(T _ {\text{c}} - T) $, the $H _ {\text{c}} $ can be made
arbitrary small by choosing the temperature $T$ close enough to
the critical temperature.

Considering the switches for current it is necessary to note a few
earlier papers~\cite{Clinton97,Oh97} on similar devices with
\underline {\emph{one}} channel of data recording that operate on
the basis of transition between superconducting and normal states.
A model of a superconducting switch device has been proposed in
Ref.~\onlinecite{Clinton97} on the basis of the F/I/S structure (I
is an insulator) in which the magnetic fringe field of a
ferromagnetic film arising due to special switch geometry is used
to control the critical current in an underlying superconducting
film.

The switch device on the basis of \textit{three-layered}
F$'$/F$''$/S structure in which the direction of the magnetization
in the relatively thin internal F$''$ layer is changed by a weak
magnetic field was theoretically studied in
Ref.~\onlinecite{Oh97}. As the mutual ordering of the
magnetizations $\mathbf{M'} $ and $\mathbf{M''} $ changes from an
antiparallel arrangement to a parallel one, the device undergoes a
transition from the \textsf{S} state to the \textsf{N} one.

We would like also to note that an experimental attempt was made
to observe the ``spin switch'' effect in the
\textit{three-layered} CuNi/Nb/CuNi system~\cite{Gu02}. Authors
succeeded in showing that the critical transition temperature is
higher for the {\sf AFMS} state than it is for the {\sf FMS}
state. However, due to a non-optimal choice of parameters of the
system the measured difference between $T _ {\text{c}} ({\sf
AFMS})$ and $T _ {\text{c}}({\sf FMS})$ did not exceed 0.005\,K.
Despite this small difference the experimental setup allows to
clearly identify the two states.

The \textit{multilayered} F/S systems in which there is an
additional competition between the 0 and $\pi $ phase types of
superconductivity have much greater potential in fundamental
studies and future device applications. In fact, in the F/S
superlattices the pair amplitude $F$ should satisfy to periodical
conditions
% $F(z+L,\omega ,I)=e^{i\varphi }F(z,\omega ,Ie^{i\chi})$,
 $F(z+L,I)=e^{i\varphi }F(z,Ie^{i\chi})$,
where $L=d_{s}+d_{f}$ is the superlattice period, and $\varphi $
and $\chi $ are the phases of superconducting and magnetic order
parameters, respectively. As it follows from the detailed analysis
carried out in works~\cite{OurUFN02,OurJETPL01,OurPRB01,OurSUST02}
on the basis of the theory of the proximity
effect~\cite{OurPRB,OurJETP_L,OurJETP,OurJETPL00} for the case of
a \textit{contact} between the dirty S and F metals, the
superconducting states of the F/S \textit{superlattice} can be
described using four different sets $\varphi \chi$: 00, $\pi$0,
0$\pi$, and $\pi \pi$. This leads to a considerably greater number
of combinations of the magnetic (\textsf{FM} or \textsf{AFM}) and
conducting (\textsf{S} or \textsf{N}) properties of the F/S
superlattices (up to five different ones) in comparison with the
\textit{trilayer} case, in which only two states (\textsf{AFMS}
and \textsf{FMN}) have been considered~\cite{LTag99,Buz_Ved}.
Hence, there is a much larger variety of regimes at which the
control devices on the basis of the F/S \textit{superlattice} can
operate. The conditions of the \textit{separate} control of the
magnetic and superconducting channels of data recording were
determined. Note that the critical thickness of the S layers $d _
{\text{s}} ^ {\text{c}} $, at which $T _ {\text{c}} $ vanishes, is
always less for the {\sf AFMS} state than for the {\sf FMS} state.
More recently, \citet{FominovJETPL03} have shown  for the F/S/F
trilayer that superconductivity has merely the {\sf AFMS} nature
at any thickness $d _ {\text{s}} > d _ {\text{s}} ^ {c} $.

Thus, as we can see, the F/S superlattices possess a number of
\textit {theoretical advantages} in comparison with the
three-layered structures, that make them a better choice for a
development of a conceptual scheme of the proposed control devices
to be used for data storage and processing. However, from a
\textit{practical} point of view in case of the superlattices it
is quite difficult to control orientation of the magnetization
separately of each F layer by an external magnetic
field~\footnote{ The control of ordering of all magnetizations
results in the need of using additional sequences of the
switching-off fields that return the system to its initial state
if the localized spins ordering in each F layer is not of an
``easy-plane'' type.}. The simplest F/S system that assumes a
competition between the 0 and $ \pi $ phase states both in
magnetism and superconductivity is a four-layered F/S/F$'$/S$'$
system with two superconducting layers and two ferromagnetic
layers that differ in their boundary conditions for the inner and
outer layers  in contrast to the F/S \emph{superlattice}
approach~\cite{OurUFN02,OurSUST02}, where the four-layered F/S/F/S
system was considered only as an \emph{elementary cell} with
\emph{periodic} boundary condition.

\section{Four-layered F/S/F$'$/S$'$ structure}

\begin{figure} [htbp]
\centerline {\includegraphics [scale=.8]{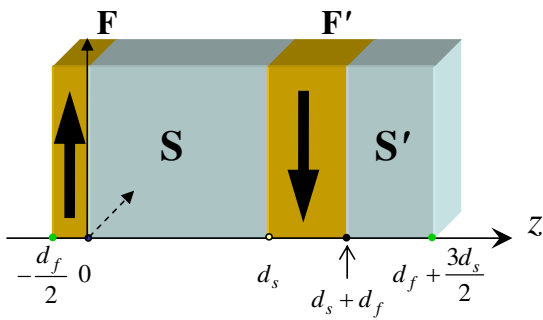}}
% \centerline{\includegraphics [scale=.4]{Proshin_EPS1.eps}}
%\centerline {\includegraphics [scale=0.7]{4layers_gray.pdf}}
%width=3.30in,height=1.93in]
\caption{(Color online) The geometry of the studied four-layered
F/S/F$'$/S$'$ system. Vertical arrows show the directions
%(in plane perpendicular to the $z$-axis)
of the magnetizations that play the role of the magnetic order
parameter (they are in a plane perpendicular to the $z$-axis).}
\label{fig1}
\end{figure}

Consider the four-layered F/S/F$'$/S$'$ system with alternating
layers along the $z$ axis (see Fig.~\ref{fig1}). Assume that both
outer F and S$'$ layers have thicknesses which are half the
thicknesses of the corresponding inner F$'$ and S layers,
$d_{\text{f}}$/2 and $d_{\text{s}}$/2, respectively. This would
allow us to simplify the solutions, and, thereafter, to compare
the obtained results for the four-layered structure with the ones
for the trilayer and superlattice cases.

The choice of this particular system allows one to take into
account a possible phase change of the superconducting and
magnetic order parameters while traversing through the F or S
layers, and to investigate mutual accommodation of the competing
BCS and LOFF types of electron pairing on the one hand, and of
superconductivity and magnetism on the other. For simplicity we
will use the 1D model when both order parameters and the pair
amplitude depend only on $z$. An extension of the results to the
3D case is straightforward (see, for example,
Refs.~\onlinecite{OurUFN02} and \onlinecite{OurUFN03}). We note
here that the distinction between solutions obtained within the
framework of the 1D and 3D models is marginal for the choice of
parameters studied below~\cite{OurJETPL01,OurPRB01}.

To find the critical temperature we assume the usual relation
between the energy parameters of the system
$\varepsilon_{\text{F}} \gg 2I \gg T_{\text{cs}}$, where
$\varepsilon_{\text{F}}$ is the Fermi energy and  $T_{\text{cs}}$
is the critical temperature of the S material. We also suppose the
dirty limit conditions
\begin{equation*}
    l_{\text{s}} \ll \xi_ {\text{s}} \ll  \xi_ {\text{s}0}, \qquad
    l_{\text{f}} \ll a_{\text{f}} \ll \xi_{\text{f}}.
\end{equation*}
\noindent
 Here $l_{\text{s,f}}=v_{\text{s,f}}\tau _{\text{s,f}}$ is the
mean free path length for the S(F) layer; $v_{\text{s,f}}$ is the
Fermi velocity; $\xi _{\text{s,f}} =(D_{\text{s,f}}/2\pi
T_{\text{cs}})^{1/2}$ is the superconducting coherence length;
$a_{\text{f}} =v_{\text{f}} /2I $ is the spin stiffness length;
$\xi_{\text{s}0}$ is the BCS coherence length;
$D_{\text{s,f}}=v_{\text{s,f}}l_{\text{s,f}} /3$ is the diffusion
coefficient.

In this case, the common boundary value problem~\cite{OurJETP} for
each layer is reduced to the Gor'kov self-consistency equations
for the ``pair amplitudes''~\footnote {Usually the term ``pair
amplitude" is used for $F(z)=\Delta(z)/\lambda(z)$. The latter in
turn is proportional to the sum of Gor'kov function $F(z,\omega)$
over Matsubara frequency $\omega$. However, in this paper we often
use this term for $F(z,\omega)$ due to
 traditions~\cite{OurUFN02} and for simplicity. } $F(z,\omega)$
\begin{equation}
\label{eq3}
\begin{array} {l}
\Delta _{\text{s}}(z)  = 2\lambda _{\text{s}} \pi T \Re
\sum\limits
_{\omega> 0} {^ {'} F_{\text{s}} (z, \omega)}, \quad \\
\Delta _{\text{f}}(z)  = 2\lambda _{\text{f}} \pi T \Re
\sum\limits _ {\omega> 0} {^ {'} F_{\text{f}} (z, \omega)}
\end{array}
\end{equation}
\noindent
 and to the Usadel equations, that appear for the S and F
 layers as follows
\begin{equation}
\label{eq1}
\begin{array} {l}
\left [{\omega - \dfrac {D_{\text{s}}} {2} \dfrac {\partial ^2}
{\partial^2 z}} \right] F_{\text{s}} (z, \omega) = \Delta _{\text{s}} (z), \\
\left [{\omega + iI - \dfrac {D_{\text{f}}} {2} \dfrac {\partial
^2}
{\partial^2 z}} \right] F_{\text{f}} (z, \omega) = \Delta _{\text{f}} (z). \\
\end{array}
\end{equation}
\noindent
 In Eqs.~(\ref{eq3}),(\ref{eq1})  $ \omega = \pi T(2n +1)$ is the
Matsubara frequency; $ \Delta _ {\text{s(f)}} $ and $ \lambda _
{\text{s(f)}} $ are the superconducting order parameter and the
electron-electron coupling constant in the S(F) layers,
respectively. The prime on the summation sign indicates cutoff at
the Debye frequency $\omega_\textrm{D}$. The diffusion coefficient
$ D_{\text{f}} $ in the F layer is assumed to be real rather than
complex~\cite{OurUFN02} since the difference between its two
values is insignificant under the conditions $2I \tau _ {\text{f}}
\ll 1$ used below (see discussion in Ref.~\onlinecite{OurUFN03}).

The coupling between the superconducting and ferromagnetic layers
is provided by corresponding  boundary conditions, which connect
the pair amplitude fluxes with the pair amplitude jumps on the
interfaces of the layers, and are written in the following
form~\cite{OurPRB,OurJETP_L,OurJETP}
 \begin{subequations}
\label{eq2tot}
\begin{equation}
\label{eq2a}
\begin{array} {c}
 \dfrac {4} {\sigma _{\text{s}} v_{\text{s}}} D_{\text{s}} \left.
 {\dfrac {\partial F_{\text{s}} (z, \omega)} {\partial z}} \right | _ {z = z_i}
 = \dfrac {4} {\sigma _{\text{f}} v_{\text{f}}} D_{\text{f}} \left. {\dfrac {\partial F_{\text{f}} (z, \omega)}
 {\partial z}} \right | _ {z = z_i} = \\
 = \pm \left [{F_{\text{s}} (z_i \pm 0, \omega) - F_{\text{f}} (z_i \mp 0, \omega)} \right]. \\
 \end{array}
\end{equation}
\noindent
  Here index  $i $ numbers the interfaces, and $z_i $
 takes the following values: $z _ {1} =0$, $z _ {2} = d _ {\text{s}}
$, $z _ {3} = d _ {\text{s}} + d _ {\text{f}} $. The upper signs
are chosen at $i=1,3$, the lower ones are chosen at $i=2$. The
pair amplitude flux through the outside boundaries ($z _ {0}  = -d
_ {f } /2$, $z _ {4} = 3d _ {\text{s}} /2 + d _ {\text{f}}$) is
absent
\begin{equation}
\label{eq2b}  \left. {\dfrac {\partial F_{\text{f}} (z, \omega)}
{\partial z}} \right | _ {z = z_0} = \left. {\dfrac {\partial
F_{\text{s}} (z, \omega)} {\partial z}} \right | _ {z = z_4} = 0.
\end{equation}

\end{subequations}
\noindent
 The last conditions~(\ref{eq2b}) distinguish the F/S/F$'$/S$'$
 case from the F/S superlattice case~\cite{OurUFN02,OurSUST02} in
 which the periodical boundary conditions are imposed.
In Eq.~(\ref{eq2a}) $ \sigma _ {\text{s(f)}}$ is the boundary
transparency at the S(F) side ($0 \le \sigma _{\text{s,f}} <
\infty$)~\cite{OurUFN02,OurJETP}. They satisfy the detailed
balance condition: $\sigma  _ {\text{s}} v _ {\text{s}} N _
{\text{s}} = \sigma  _ {\text{f}} v _ {\text{f}} N _ {\text{f}} $,
where $N _ {\text{s(f)}} $ is the density of states at the Fermi
level.

In order to calculate the critical temperatures of this F/S system
taking into consideration the boundary transparencies, thicknesses
of layers, etc. we should solve the system of equations
(\ref{eq1}) and (\ref{eq2tot}) together with the self-consistency
equations~(\ref{eq3}).

The powerful pair-breaking action of the exchange
field $I $ ($I \gg \pi T _ {\text{cs}} $) is the
basic mechanism for the destruction of
superconductivity  in the F/S systems. For
simplicity~\cite{OurJETP} assume that $ \lambda _
{\text{f}} = 0$ ($ \Delta _ {\text{f}} =0$) in the F
layers. We will search the solutions of equations
(\ref{eq3})-(\ref{eq2tot}) for the inner layers as a
linear combination of symmetric and antisymmetric
functions relative to the centers of the S and F$'$
layers. The pair amplitudes look the same as in the
superlattice case~\cite{OurPRB01}. The zero flux of
the pair amplitude through the outside
boundaries~(\ref{eq2b}) determines only the even
cosine-like functions for the outer layers. At these
boundaries the antinodes should be fixed. Thus the
simplest solutions of the boundary value problem for
the F/S/F$'$/S$'$ system have the following form
\begin{widetext}
\begin{subequations}
\label{eq4ac}
\begin{eqnarray}
F_{\text{f}} &=& B\dfrac {\cos k_{\text{f}} (z + d_{\text{f}} /
2)} {\cos (k_{\text{f}} d_{\text{f}} / 2)}, \qquad  - d _
{\text{f}} /2  < z  <0,
 \label{eq4a}\\ %(4a)
F_{\text{s}} &=& A\dfrac {\cos k_{\text{s}} (z - d_{\text{s}} /
2)} {\cos (k_{\text{s}} d_{\text{s}} / 2)} + C\dfrac {\sin
k_{\text{s}} (z - d_{\text{s}} / 2)} {\sin (k_{\text{s}}
d_{\text{s}} / 2)}, \qquad
 0 <z < d _ {\text{s}},
 \label{eq4b}\\ %(4b)
F_{\text{f}}' &=& B'\dfrac {\cos k_{\text{f}}' (z - d_{\text{s}} -
d_{\text{f}} / 2)} {\cos (k_{\text{f}}' d_{\text{f}} / 2)} +
D'\dfrac {\sin k_{\text{f}}' (z - d_{\text{s}} - d_{\text{f}} /
2)} {\sin (k_{\text{f}}' d_{\text{f}} / 2)}, \qquad  d _
{\text{s}} <z <(d _ {\text{f}} +d _ {\text{s}}),
 \label{eq4c}\\  %(4s)
F_{\text{s}}' &=& A'\dfrac {\cos k_{\text{s}}' (z - d_{\text{f}} -
3d_{\text{s}} / 2)} {\cos (k' _ s d_{\text{s}} / 2)}, \qquad  (d _
{\text{f}} + d _ {\text{s}}) < z < (d _ {\text{f}} + 3d _
{\text{s}} /2).
 \label{eq4d} %(4d)
\end{eqnarray}
\end{subequations}
\end{widetext}
\noindent
 Here $k _ {\text{s(f)}} $ and $k' _ {\text{s(f)}} $ are
the components of the wave vector that describe spatial changes of
the pair amplitudes $F_{s(f)}  $ and $F'_{s(f)}  $ across the
layers (along the $z$ axis) independently of the frequency
$\omega$. The chosen form of the $F_{\text{s}}$ and
$F_{\text{f}}'$ pair amplitudes is related to the symmetry of
F/S/F$'$/S$'$ system. In Eqs.~(\ref{eq4b}) and (\ref{eq4c}) the
first terms are responsible for the symmetric superconducting 0
phase solutions, while the second terms are responsible for the
appearance of superconducting antisymmetric $\pi$ phase solutions
(see below discussion in Sec. IV).

Since we are mainly interested in performing qualitative studies
of the properties of the F/S/F$'$/S$'$ nanostructure, the
single-mode approximation~(\ref{eq4ac}) is used to obtain the
analytical solution of the complicated boundary value problem.
However, when quantitative estimates are needed (to fit
theoretical results to experimental data) the latter approximation
works well only for a certain range of the values of the
parameters in the problem~\cite{FominovJETPL01,FominovPRB02}.
According to our estimates~\cite{OurJETP} the optimal set of
parameters used below is close to this range ($d_{\text{s(f)}}>\xi
_{\text{s(f)}}$). Note that in any approximation (single-mode,
multi-mode, etc~\cite{FominovPRB02}) the symmetry of the problem
solutions will be different for the different S and S$'$ (F and
F$'$) layers. This fact reflects the general property of the
studied system: the nonequivalence of layers of the same type,
which results in different superconducting properties of the
internal S and outer S$'$ layers.

Substituting the solutions (\ref{eq4ac}) into the self-consistency
equations (\ref{eq3}) and performing the standard summation over
 $\omega$ we derive the usual Abrikosov-Gor'kov type equation
for the reduced superconducting transition temperatures
$t_{\text{c}},t'_{\text{c}}$ of the S and S$'$ layers,
respectively
\begin{equation}
\label{eq5} \ln t_c = \Psi \left ({\dfrac {1} {2}} \right) - \Re
\Psi \left ({\dfrac {1} {2} + \dfrac {D_{\text{s}} k_{\text{s}}
^2} {4\pi T _ {\text{cs}} t_c}} \right)
\end{equation}
 \noindent
  where $t_{\text{c}}=T_c/T_{cs}$; $ \Psi (x) $ is the digamma
function, and the pair-breaking parameter $D _ {\text{s}} k _
{\text{s}} ^ {2} $ is the solution of the other transcendental
equation (see Eqs.~(\ref{eq9})-(\ref{eq10}), and (\ref{eq7})
below), which may differ not only for each of the possible phases,
but also for each superconducting layer (S and S$'$) as well. To
get an equation for $t'_{\text{c}}$ it is necessary to exchange
$t_{\text{c}}$ for $t'_{\text{c}}=T'_c/T_{c,s}$ and $k_s$ for
$k'_s$ in Eq.~\ref{eq5} (see also Eq.~(\ref{eq7}) and its
discussion below).

Substituting (\ref{eq4ac}) in (\ref{eq2tot}) we obtain a set of 6
equations for factors $A$, $B$, $C$, $A'$, $B'$, $D'$
\begin{equation}
\label{eq6} \left\{
\begin{array}{ll}
B + \alpha A + \beta C = 0, \quad %\\
 & \gamma B - A + C = 0, \\
 \alpha A - \beta C + {B} ' - {D} ' = 0, \quad\quad % \\
 & - A - C + {\gamma} ' {B} ' - {\delta} ' {D} ' = 0,  \\
 {B} ' + {D} ' + {\alpha} ' {A} ' = 0,  \quad\quad %\\
& {\gamma} ' {B} ' + {\delta} ' {D} ' - {A} ' = 0, \\
 \end{array}
 \right.
\end{equation}
 \noindent where following notations are introduced
\begin{equation}
\label{eq7}
\begin{array} {lcllcl}
 \alpha & = & \dfrac {4D_{\text{s}} k_{\text{s}}} {\sigma _{\text{s}} \upsilon _{\text{s}}} \tan\dfrac {k_{\text{s}} d_{\text{s}}} {2} - 1, \quad\quad%\\
& \beta & = & \dfrac {4D_{\text{s}} k_{\text{s}}} {\sigma _{\text{s}} \upsilon _{\text{s}}} \cot\dfrac {k_{\text{s}} d_{\text{s}}} {2} + 1, \\
 \gamma & = & - \dfrac {4D_{\text{f}} k_{\text{f}}} {\sigma _{\text{f}} \upsilon _{\text{f}}} \tan\dfrac {k_{\text{f}} d_{\text{f}}} {2} + 1, \quad\quad%\\
& \delta & = & \dfrac {4D_{\text{f}} k_{\text{f}}} {\sigma _{\text{f}} \upsilon _{\text{f}}} \cot\dfrac {k_{\text{f}} d_{\text{f}}} {2} + 1. \\
 \end{array}
\end{equation}
 \noindent
Quantities related to the outer superconducting S$'$ layer ($
\alpha '$) or the inner F$'$ layer ($ \gamma '$, $ \delta '$) are
marked with \textit{a prime}. The prime also appears at
corresponding wave vectors ($k _ {\text{s}}' $ or $k _ {\text{f}}'
$) on the right side of the expressions (\ref{eq7}). In the
framework of the made approximations the complex value of the wave
vectors is defined as follows
 %\begin{equation}
\begin{subequations}
\label{eq8}
\begin{eqnarray} %{lll}
 k_{\text{f}}^2 &=& \left ({{k} ' _{\text{f}}} \right) ^2 = - \dfrac {2iI}
 {D_{\text{f}}};
 \label{eq8a}  \\
 k_{\text{f}}^2 &=& - \dfrac {2iI} {D_{\text{f}}}, \quad \left ({{k} ' _{\text{f}}} \right) ^2
 = \dfrac {2iI} {D_{\text{f}}} = \left ({k_{\text{f}}^2} \right) ^ \ast .
  \label{eq8b}  %\\
 \end{eqnarray}
 \end{subequations}
%\end{equation}
 \noindent The equations~(\ref{eq8a}) are valid for the case of the mutual ferromagnetic
ordering of the magnetizations in the F and F$'$ layers, and the
equations~(\ref{eq8b}) are valid for the case of the
antiferromagnetic ordering when  $I' = - I $, , i.e. the phase
$\chi$ of the magnetic order parameter equals $ \pi $.

Note that within the 3D model~\cite{OurUFN03,OurUFN02,OurJETPL00}
$k_f^2$ and $(k'_f)^2$ in Eqs.~\ref{eq8} should be replaced by
$(k_f)^2+q_f^2$ and $(k'_f)^2 + q_f^2$, respectively. Here $q_f$
is a wave vector in the F/S boundary plane, which is responsible
for the 2D interface LOFF states with the spatial oscillations of
the pair amplitude in the $x-y$ plane. The concrete value of $q_f$
is found from the condition for maximum of $T_c$. However, in our
case the difference between the 1D ($q_f \equiv 0$) and 3D
($q_f\neq 0$) approaches is unessential for the selected set of
the system parameters (see Section~IV below).

Thus, the expressions obtained above include a
competition between the 0 phase and the $ \pi $
phase types of superconductivity. They also take
into account interaction of the localized moments of
layers F and F$'$ through the superconducting layer
S. The quantum coupling both between the adjacent S
and S$'$ layers through the F$'$ layer and between
the adjacent F and F$'$ layers through the S layer
is provided by superconducting correlations of
conduction electrons. It is known~\cite{de_Gennes}
that the role of the \textit{true} superconducting
order parameter for the heterogeneous systems
discussed in the paper is played by a pair amplitude
$F(z) = \Delta(z)/\lambda(z)$. In contrast to a
parameter $\Delta(z)$, the pair amplitude does not
vanish in the ferromagnetic layer, but provides a
quantum coupling between the layers via
superconducting correlations which have an
inhomogeneous oscillatory behavior due to the
LOFF-like type of pairing.

The pair-breaking parameters  $D _ {\text{s}} k _ {\text{s}} ^ {2}
$ for the S layer and $D _ {\text{s}} (k _ {\text{s}}') ^ {2} $
for the S$'$ layer should be determined from the condition of
nontrivial compatibility of the set of equations~(\ref{eq6}). It
is possible to factorize the corresponding determinant and to
obtain the following equation
 \begin{equation}
\label{eq75}
\begin{array} {l}
  \left( {\alpha '\delta ' + 1} \right)
  \left[ {\left( {\alpha \gamma  + 1} \right)
  \left( {\beta \gamma ' - 1} \right) +
  \left( {\beta \gamma  - 1} \right)\left( {\alpha \gamma ' + 1} \right)} \right] + \\
   + \left( {\alpha '\gamma ' + 1} \right)
   \left[ {\left( {\alpha \gamma  + 1} \right)
   \left( {\beta \delta ' - 1} \right)
   + \left( {\beta \gamma  - 1} \right)
   \left( {\alpha \delta ' + 1} \right)} \right] = 0.
\end{array}
\end{equation}
 \noindent
Equation~(\ref{eq75}) can be simplified by taking into account the
independence of the solutions for the S and S$'$ layers and
knowing the solutions for the superlattice case. It is possible to
obtain the following sets of equations for $k _ {\text{s}} $ and
$k _ {\text{s}}'$, which are different for the {\sf FM} and {\sf
AFM} configurations.  Note that only equations leading to the
finite nonzero critical temperature are kept in these sets (see
Eqs.~(\ref{eq9}) and (\ref{eq10}) below).

For the {\sf FM} ordering of the magnetizations we obtain two
cases {\sf FM}($a$) and {\sf FM}($b$)
 \begin{equation}
\label{eq9}
\begin{array} {l}
 \text {{\sf FM}}
 \left ( {\begin{array} {l} a\\ a' \end{array} } \right )
 \Rightarrow
 \left \{ { \begin{array} {ll}
 \alpha \gamma + 1 = 0\; \leftarrow \text{layer S} \\
 {\alpha} '\gamma + 1 = 0\; \leftarrow \text{layer S}' \\
 \end{array}} \right. \\
 \text {{\sf FM}}
 \left ( {\begin{array} {l} b\\b' \end{array} } \right )
 \Rightarrow
 \left \{ {
 \begin{array}  {ll}
 2\alpha \gamma \beta \delta + \left ({\beta - \alpha} \right) \left
 ({\gamma + \delta} \right) = 2\; \leftarrow \text {layer S} \\
{\alpha} '\delta + 1 = 0\; \leftarrow \text {layer S}'. \\
 \end{array}} \right. \\
 \end{array}
\end{equation}
 \noindent Here $\gamma '$ and $\delta '$ for the F$'$ layer are
substituted by $\gamma $ and $\delta $, respectively, due to
Eq.~(\ref{eq8a}).

According to equation~(\ref{eq8b}), when $k _
{\text{f}} ' =k _ {\text{f}} ^ {\ast} $ (leading to
$\gamma ' =\gamma ^ {\ast}$ and $\delta ' =\delta ^
{\ast}$), for the {\sf AFM} ordering we have two
other cases {\sf AFM}($c$) and {\sf AFM}($d$)
 \begin{equation}
\label{eq10}
\begin{array} {ll}
 \text{{\sf AFM}}
 \left ( {\begin{array} {l} c\\c' \end{array} } \right )
 \Rightarrow
  \left \{
 {\begin{array} {l}
 \alpha \beta \left | \gamma \right | ^ 2 + \left ({\beta - \alpha} \right) \Re \gamma = 1 \; \leftarrow \text{layer S} \\
 {\alpha} '\gamma ^ \ast + 1 = 0\; \leftarrow \text {layer S}' \\
 \end{array}} \right. \\
 \text {{\sf AFM}}
 \left ( {\begin{array} {l} d\\d' \end{array} } \right )
 \Rightarrow
  \left \{
 {\begin{array} {l}
 2\alpha \gamma \beta \delta ^ \ast + \left ({\beta - \alpha} \right) \left ({\gamma + \delta ^ \ast} \right) = 2\; \leftarrow \text{layer S} \\
 {\alpha} '\delta ^ \ast + 1 = 0\; \leftarrow \text {layer S}'. \\
 \end{array}} \right.  \\
 \end{array}
\end{equation}

In the general case, there are 4 different solution sets {\sf
FM}($a,a'$), {\sf FM}($b,b'$), {\sf AFM}($c,c'$), and {\sf
AFM}($d,d'$) for the S and S$'$ layers, each of which completely
defines the state of \textit{both} layers and, hence, the
corresponding reduced transition temperatures $t_{\text{c}}$ and
$t'_{\text{c}}$ (\ref{eq5}). However, since the solution
$t_{\text{c}}$ of Eq.~(\ref{eq5}) does not change when $k_s$ is
replaced by its complex conjugate the solutions $t'_{\text{c}}$
for the S$'$ layer do not depend on relative orientation of the
magnetizations: the solution of Eq.~(\ref{eq5}) for the {\sf
FM}($a'$) case coincides with the solution for the {\sf AFM}($c'$)
case. The same is true for the solutions for the {\sf FM}($b'$)
and {\sf AFM}($d'$) cases.

Moreover, two different solutions for the outer S$'$ layer ({\sf
FM}($a'$)$=${\sf AFM}($c'$) and {\sf FM}($b'$)$=${\sf AFM}($d'$))
always coincide with the solutions for the F/S
superlattice~\cite{OurUFN02,OurJETPL01,OurPRB01}, in which an
orientation of the magnetizations is common to all F layers (the
{\sf FM} case). We try to classify the
solutions~(\ref{eq9}),(\ref{eq10}) for the F/S/F$'$/S$'$ system
following the classification scheme proposed for a
\textit{superlattice} in
Refs.~\onlinecite{OurUFN02,OurJETPL01,OurPRB01} (see also
Section~II).  As it follows from that classification there are
four possible states for a \textit{superlattice}, which are
described using two possible values (0 and $\pi$) for the phases
of the superconducting and magnetic order parameters. Two pairs of
equations for the $k_{\text{s}}'$ in the S$'$ layer that coincide
{\sf FM}($a'$)={\sf AFM}($c'$) and {\sf FM}($b'$)={\sf AFM}($d'$)
lead to the 00 and $ \pi 0$ solutions, respectively: the first
symbol corresponds to the superconducting order parameter phase
($\varphi$), the latter one corresponds to the magnetic order
parameter phase ($\chi$). Thus, we have only two distinguishable
solutions $a'$ and $b'$ for the S$'$ layer (see below
Fig.~\ref{fig2}a,b).

The latter can be easily understood from the physical point of
view. Only \textit{one} ferromagnetic layer (F$'$) acts on the
outer S$'$ layer. As a result the state of the layer depends only
on the magnitude of the exchange field in the F$'$ layer and does
not depend neither on its sign nor on mutual ordering of the
magnetizations. In other words, the S$'$ layer is
\textit{\textbf{always}} in the \textbf{\textit{local
ferromagnet}} (\textsf{FM}) environment, therefore the $\pi$
magnetic solutions do not exist for this layer.

For the S layer we have also two known superlattices solutions,
namely, the {\sf FM}($a$) solution, that leads to the 00 solution,
and the {\sf AFM}($c$) solution, that leads to the $0\pi $
solution (the 0 phase superconductivity and the $ \pi $ phase
magnetism). Finally, there are two \emph{extra} solutions {\sf
FM}($b$) and {\sf AFM}($d$). Their presence is related to the
external boundary conditions~(\ref{eq2b}) since the pair
amplitudes (\ref{eq4a}), (\ref{eq4d}) contain only even cosine
solutions. These states are the $\pi$ superconducting states,  and
in order to distinguish them from the earlier mentioned
\textit{superlattice} solutions we will denote these ones with
tilde $\widetilde{\pi \chi}$ ($\chi=0, \pi$ is the phase of the
magnetic order parameter). Thus the {\sf FM}($b$) solution
determines the $\widetilde{\pi0}$ state of the four-layered system
and the {\sf AFM}($d$) solution corresponds to the
$\widetilde{\pi\pi}$ one.

In the next section we will examine the obtained solutions and
clarify the winners in the interplay of the four states
(\ref{eq9}),(\ref{eq10}).

\section {Discussion of the phase diagrams}

Taking into account the notations of Eq.~(\ref{eq7}) the sets of
Eqs.~(\ref{eq5}),(\ref{eq8})-(\ref{eq10}) can be used to study the
dependence of critical temperatures ($t_{\text{c}} $ and
$t'_{\text{c}}$) of the four-layered F/S/F$'$/S$'$ system on the
reduced thicknesses of the superconducting and magnetic layers,
$d_{\text{s}}/\xi_{s0}$ and $d _ {\text{f}} / a _ {\text{f}} =
\widetilde{d}$.

\begin{figure} [tbph]
%\centerline {\includegraphics [width=3.98in, height=2.67in]
%{PROSHIN_BI_12.eps}}
\centerline {\includegraphics [scale=.5]{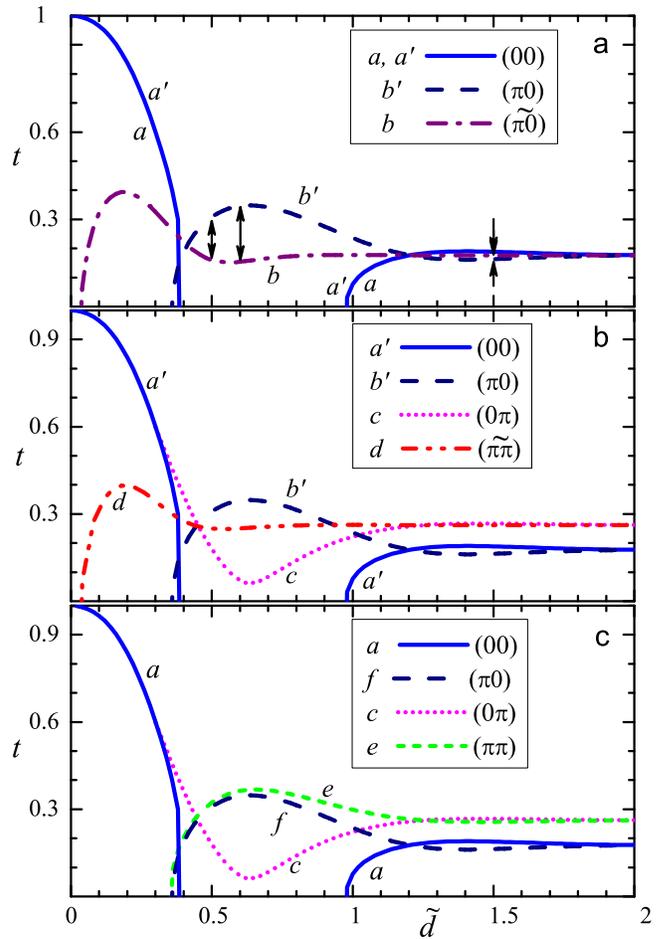}}
 \caption {(Color online) The phase diagrams ($t-\widetilde{d}$) of the F/S nanostructures
for the following values of parameters: $ \sigma _ {\text{s}}  =
15 $, $2I \tau _{\text{f}} = 0.1 $, $n _ {\text{sf}} =  1.4 $, $l
_ {\text{s}} =0.25 \xi _ {s0} $, and $d _ {\text{s }} = 0.72 \xi _
{s0}$. In the figure $t=T/T_{\text{cs}}$ is the reduced
temperature, and $\widetilde{d}=d _ {\text{f}}/a _ {\text{f}} $ is
the reduced F layer thickness. The dependences of the reduced
critical temperatures $t_{\text{c}}$ and $t'_{\text{c}}$ for the
F/S/F$'$/S$'$ system versus $\widetilde{d}=d _ {\text{f}}/a _
{\text{f}} $ are presented in panels a and b.  The $t'_{\text{c}}$
curves for the outer S${'}$ layer are denoted using letters
\textit{with a prime}. The letters \textit{without a prime}
indicate the $t_{\text{c}}$ curves for the inner S layer.
\\
(a) -- the phase diagram of the four-layered system
for the ferromagnetic (\textsf{FM}) configuration of
the magnetizations of both F layers. The arrows show
the $(t_{\text{c}}-t'_{\text{c}})$
difference between the states which are discussed in the paper. \\
(b) -- the phase diagram of the four-layered system
for the \textsf{AFM} configurations.\\
(c) -- the phase diagram of the F/S superlattice
where the thicknesses of all F layers equal
$d_{\text{f}}$, and the thicknesses of all S layers
equal $d_{\text{s}}$. In this case all S layers have
the same critical temperature. That is why only the
$t_{\text{c}}(\widetilde{d})$ curves are shown here.
The $a$ and $c$ curves form the phase diagram for
the F/S/F trilayer in which the thicknesses of both
F layers are equal to $d _{\text{f}}/2 $. }
 \label{fig2}
\end{figure}

There are four more theoretical parameters of the system ($\sigma_
{\text{s}},2I\tau _ {\text{f}},n _ {\text{sf}},l _ {\text{s}}$),
that are necessary to consider in a general case. Keeping in mind
a possibility of an application of the system as a  ``control
device" we have searched for such a set of parameters for which
the difference between the various states of the F/S/F$'$/S$'$
system is sufficiently large to be observed. After performing
numerous computer experiments we have found a range for the values
of the parameters that satisfies these conditions. The
\textit{optimal range of parameters} should be as follows: the
boundary should be sufficiently transparent ($\sigma_
{\text{s}}\gtrsim 5 \gg 1 $), the ferromagnetic metal should be
sufficiently dirty or (and) weak enough in regard to its magnetic
properties ($2I\tau _ {\text{f}} = l _ {\text{f}}/a _ {\text{f}}
\lesssim 0.15 \ll 1$), and the parameter $n _ {\text{sf}} = N _
{\text{s}} v _ {\text{s}} / N _ {\text{f}} v _ {\text{f}} > 1$.
From experimental viewpoint this constraint on the values of the
parameters does not look unreasonable. Note, the stronger is an
implementation of each inequality, the larger can be a difference
between critical temperatures for the 0 and $\pi$ magnetic states.
The choice of any one of these parameters lying out-of-range leads
to significant decreasing that difference. Parameters
$d_{\text{s}}$ and $l_{\text{s}}$ should satisfy to the ``dirty
limit" conditions. Their influence on the noted above difference
is minimal. Though all values are important for a shape of the
$T_c(d_{\text{f}})$ dependence (this has been detailed in the
recent review~\cite{OurUFN02}).

A set of phase curves $t_{\text{c}} (d _ {\text{f}}) $ and
$t'_{\text{c}} (d _ {\text{f}}) $ for the \textit{optimal} values
of the parameters is shown in Fig.~\ref{fig2}. The notation used
for the curves corresponds to the notation used in
Eqs.~(\ref{eq9}),(\ref{eq10}). The curves $c'$ and $d'$ are not
shown in Fig.~\ref{fig2} since $c'\equiv a'$ and $d'\equiv b'$.

As one might expect, the $a' $ and $b' $ curves for the S$'$ layer
in Figs.~\ref{fig2}a and \ref{fig2}b are identical for both
\textsf{FM} and \textsf{AFM} configurations, and the $a' $ curve
for the S$'$ layer coincides completely with the $a $ curve for
the S layer since both of them describe the same 00 state
according to the superlattice classification
scheme~\cite{OurJETPL01,OurPRB01,OurUFN02}. The rest of the states
for the inner S layer (the $b$ curve for the \textsf{FM}
configuration, and the $c$ and $d$ curves for the \textsf{AFM}
one) have different dependencies as compared with the ones for the
S$'$ layer and each other.

The phase diagram for the F/S \textit{superlattice}
(Fig.~\ref{fig2}c) is obtained following procedures developed in
Refs.~\onlinecite{OurUFN02,OurJETPL01,OurPRB01}. Contrary to the
\textit{four-layered} F/S/F$'$/S$'$ system in the
\textit{superlattice} case the different S layers have the same
critical temperatures due to the periodicity condition imposed on
the pair amplitudes.
%~\cite{OurUFN02,OurJETPL01,OurPRB01}.
%In this figure the known solutions $00$, $\pi 0$, $0\pi$, $\pi
%\pi$ are represented.
The $a$, $c$, and $f$ curves for the F/S superlattice completely
coincide with the $a$, $c$, and $b'$ ones for the four-layered
system, respectively. The $\pi \pi$ superlattice state (the $e$
curve), which is not found for our four-layered system, can be
obtained from Eq.~(\ref{eq5}) where corresponding parameter $D_s
k_s^2$ is determined by following equation
$$
\alpha \beta \left | \delta \right | ^ 2 + \left ({\beta - \alpha}
\right) \Re \delta = 1.
$$
To obtain the phase diagram for the F/S/F trilayer
we have to exclude the states with the $\pi$ phase
superconductivity (the $f$ and $e$ curves) from this
figure. Hence as it follows from Fig.~\ref{fig2} and
discussions presented here the four-layered system
has more physically different states than the F/S/F
trilayer and even the F/S superlattice!

Thus, the $00$, $\pi0$, and $0\pi$ states of the four-layered
system (Figs.~\ref{fig2}a,b) correspond to the same states of the
superlattice (Fig.~\ref{fig2}c). The $\widetilde{\pi 0}$ state
(the $b$ curve in Fig.~\ref{fig2}a) and the $\widetilde{\pi \pi}$
one (the $d$ curve in Fig.~\ref{fig2}b) are the states associated
with the $\pi$ phase superconductivity. These are two extra
solutions which are not found in the superlattice case
(Fig.~\ref{fig2}c). The main difference between these
$\widetilde{\pi \chi}$ and the known $\pi\chi$ states is the peak
position. For the inner S layer it is shifted to the lower values
of the $d_{\text{f}}$ thickness compared with the superlattice
case due to the implementation of the external boundary conditions
(\ref{eq2b}).

The above-mentioned peculiarities of the
four-layered system lead to \textbf{\emph{different
critical temperatures}} of different S layers. To
show this consider the \textsf{FM} configuration
(Fig.~\ref{fig2}a) in detail. If there is no
difference between $t_{\text{c}}$ and
$t'_{\text{c}}$ for the 00 state then the case of
the $\pi$ phase superconductivity is more
interesting since there is a difference between
$t'_{\text{c}}(\pi0)$ and
$t_{\text{c}}(\widetilde{\pi0})$. Actually for each
superconducting layer the upper envelope curve is
realized due to free energy minimum condition. In
the case of the \textsf{FM} configuration that will
be $a'-b'-a'$ curve and $a-b-a$ one for the S$'$ and
S layers, respectively. This leads to switching the
ground state between the states  with the 0 and
$\pi$ superconducting phases as the thickness
$\widetilde{d}$ changes (at $\widetilde{d} \sim 0.4$
and $\widetilde{d} \sim 1.2$, respectively).

In the $\pi$ phase superconductivity case, the order
parameter $\Delta$ has opposite signs for the S and
S$'$ layers. Accordingly, the pair amplitude in the
inner F$'$ layer (Eq.~\ref{eq4c}) has a sine-like
behavior ($B'=0$) and is antisymmetric with respect
to the layer center at which the sign change of the
pair amplitude takes place while traversing the F$'$
layer. The above-mentioned different $T_c$ behavior
in the S and S$'$ layers (the $b$ and $b'$ curves in
Fig.~\ref{fig2}a, respectively) leads to a
difference between critical temperatures $t$ and
$t'$. For instance, at $\widetilde{d} =1.5$ the
reduced critical temperature of the S layer
$t_{\text{c}}$ is equal to 0.177, and $t'_{\text{c}}
= 0.163$, at $\widetilde{d}=0.5$ the difference is
larger: $t'_{\text{c}} = 0.308$ and $t_{\text{c}} =
0.16$. If the reduced thickness $\widetilde{d}$ were
equal 0.6, the difference would be almost maximal:
$t'_{\text{c}} = 0.346$, and $t_{\text{c}} = 0.154$.

The reduced critical temperatures $t_{\text{c}}$ and
$t'_{\text{c}}$ that correspond to these three values of the
reduced thickness $\widetilde{d}$ are shown in Fig.~\ref{fig2}a by
arrows. The difference between two critical temperatures
$t_{\text{c}}$ and $t'_{\text{c}}$ should be observed in
experiments with the special field-cooled samples prepared with
the \textsf{FM} ordering of the magnetizations (see
Ref.~\onlinecite{Goff02} for experimental details).

The appearance of the critical temperature difference in the
four-layered F/S system is a manifestation of the\textbf{\emph{
critical temperatures hierarchy}} in its clearest form. The origin
of the $T_{\text{c}}$ difference is obvious because, firstly, the
S and S$'$ layers are in different magnetic environment and,
secondly, they have different boundary conditions. In particular
it is expressed in the above mentioned shift of the peak of the
$\widetilde{\pi 0}$ dependence due to the outside boundary
conditions.

For the \textsf{AFM} configuration of the F/S/F$'$/S$'$ system we
have a similar picture (Fig.~\ref{fig2}b), but in this case there
are four different curves. Note that all above mentioned
peculiarities take place as well. As it has been discussed above,
the phase curves for the S$'$ layer are the same for both the
\textsf{FM} and the \textsf{AFM} orientations. Two different
solutions are obtained for the inner S layer. One of them is the
known ``superlattice" solution $0\pi$ (curve $c$) while the second
one is the $\widetilde{\pi \pi}$ solution (curve $d$). There is
also a competition between the 0 and $\pi$ phase superconductivity
that leads to a appearance of the corresponding envelope curves of
the second order phase transition for the S and S$'$ layers
($c-d-c$ and $a'-b'-a'$, respectively).

The $0\pi $ solution corresponds to the $D' = 0$ and $B' \ne 0$
case, and the $A$ and $C $ factors are not equal 0, i.e. the pair
amplitude in the S layer does not possess any parity. The
admixture of the sine solutions to the cosine ones in expression
(\ref{eq4b}) reflects the partial compensation of the paramagnetic
effect of exchange field $I $ for the S layer in the {\sf AFM}
state with antiparallel alignment of the F layers magnetizations.
The previous statement applies to the $\widetilde{\pi \pi}$ state
in the S$'$ layer too.

As in stated above \textsf{FM} case, the difference between
$t_{\text{c}}$ and $t'_{\text{c}}$ can be observed in experiments
with the special field-cooled  \textsf{AFM} samples.

\begin{figure} [htbp]

% \centerline {\includegraphics [scale=.5]{Figa5_new3.eps}}
%\centerline {\includegraphics [scale=.5]{Figa5.eps}}
\centerline {\includegraphics [scale=.43]{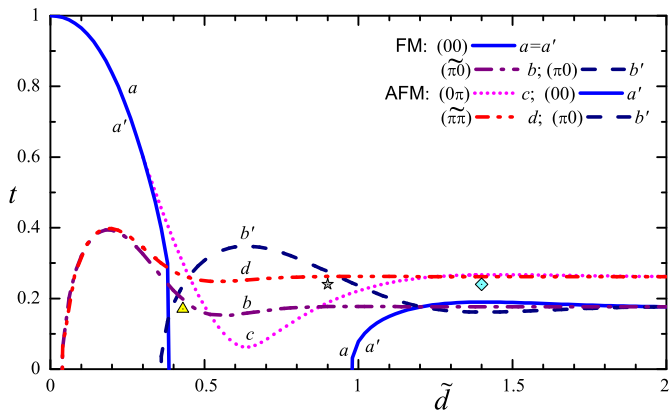}}
 \caption{(Color online) The combined phase diagram of the four-layered
F/S/F$'$/S$'$. All parameters and notations correspond to the ones
used in Fig.~\ref{fig2}. The symbols discussed in Sec.~V
correspond to the ``working" points of the proposed control device
with various number of possible states. } \label{fig5}
\end{figure}

Let us take up the common case, when there is the interplay of all
the four states (\ref{eq9}),(\ref{eq10}), to clarify the winners
in this competition. For convenience all the phase curves are
shown in Figs.~\ref{fig2}a,b in one combined diagram
(Fig.~\ref{fig5}).

Note, that at $\widetilde{d}=0.5$ the $\widetilde{\pi \pi}$ state
(the $d$ curve in Fig.~\ref{fig2}b and in Fig.~\ref{fig5}) has the
highest $T_c$ among all possible states for the S layer
$t_{\text{c}}\simeq 0.25$, but that is lower than the appropriate
temperature for the $\pi 0$ state of the S$'$ layer
$t'_{\text{c}}\simeq 0.31$. According to the theory of
second-order phase transitions, the state possessing the lower
free energy (higher $T _ {\text{c}})$ is realized. Thus for the
samples with the reduced thickness $\widetilde{d}=0.5$ the S and
S$'$ layers are both in the normal (\textsf{N}) state if the
temperature $t>t'_c\simeq 0.31$. Below $t'_c$ the S$'$ layer
becomes superconducting (\textsf{S}) but the S layer remains in
the \textsf{N} state while $t>t_c\simeq 0.25$. Finally, at $t<t_c$
the \textsf{AFMS} state (\textsf{AFM}($d,d'\equiv b'$)) wins and
for the whole system we have the case with the $\pi$ phase
superconductivity and the $\pi$ phase magnetism.

\begin{figure} [htbp]
%\centerline {\includegraphics [width=3.98in, height=2.67in]
%{PROSHIN_BI_12.eps}}
 \centerline {\includegraphics [scale=.43]{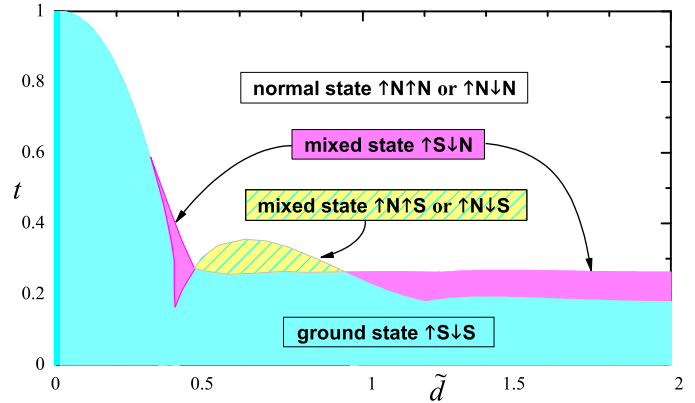}}
%\centerline {\includegraphics [scale=.5]{Figa4.eps}}
 \caption{(Color online) The generalized phase diagram of the four-layered
F/S/F$'$/S$'$ system with the same parameters as in
Figs.~\ref{fig2},\ref{fig5}. Vertical arrows show the direction of
the magnetization in the corresponding ferromagnetic layer. The
letters \textsf{S} and \textsf{N} stand for the superconducting
and normal states of the superconducting layers, respectively. For
simplicity the magnetization is assumed to be fixed and directed
``upwards'' in the outer F layer (see also the next Section).}
 \label{fig4}
\end{figure}

At $\widetilde{d}=1.5$ we have the following chain of the second
order phase transitions:
 $ \uparrow $\textsf{N}$\downarrow $\textsf{N}
 (or $ \uparrow $\textsf{N}$\downarrow $\textsf{N})
$\stackrel{t_{\text{c}}\simeq 0.27}{\longrightarrow}$
 $\uparrow $\textsf{S}$ \downarrow $\textsf{N}
$\stackrel{t_{\text{c}}'\simeq 0.19}{\longrightarrow}$
 $\uparrow$\textsf{S}$ \downarrow $\textsf{S}
(see a caption to Fig.~\ref{fig4} for an explanation of the
notation). Thus at low temperatures the \textsf{AFMS} state
(\textsf{AFM}($c,c'\equiv a'$)) wins too, but this state is
associated with the 0 phase superconductivity. Note in the
framework of our theory only transition temperatures can be found
and it is not possible to determine what state inside the
``normal'' state region is preferable.

The analogous analysis can be carried out for the entire range of
the reduced F layer thicknesses ($0<\widetilde{d}<2$). Assume the
system can choose its own state according to the theory of the
second order phase transitions. The state with higher critical
temperature wins, and one of four states defined by
Eqs.~(\ref{eq9})-(\ref{eq10}) (see also Figs.~\ref{fig2}a,b and
\ref{fig5}) is realized for the system. A complete phase diagram
constructed for the system is presented in Fig.~\ref{fig4}. Four
different regions can be defined for this diagram: at high
temperatures both S and S$'$ layers are in normal state and the
mutual ordering of the magnetizations in the F and F$'$ layers is
unimportant. As it follows from the phase diagram, there are two
regions marked in dark grey color (magenta in color online
version) with mixed antiferromagnetic state for which the inner S
layer is superconducting (\textsf{S}), and the outer S$'$ one is
normal (\textsf{N}). The striped light grey (yellow) marked region
corresponds to the mixed state with the superconducting outer S$'$
layer and the normal S layer. Finally, at low temperatures and/or
at small $d_f$ thicknesses the system is in the ground
\textsf{AFMS} state (grey (blue) marked region).

Thus, if the inner S layer is in the \textit{superconductive}
state then ordering of the magnetizations should be
\textit{antiferromagnetic}. This is the result of the
\textit{inverse} action of \textit{superconductivity on
magnetism}.

Note, the details of the phase diagram significantly depend on the
choice of the system parameters and the above analysis was carried
out assuming the absence of an external magnetic field ($H=0$).

\section{Control device scheme}

In this section we propose a conceptual scheme of a ``control
device" based on the four-layered F/S/F$'$/S$'$ structure.

Following the previous studies performed on spin
valves~\cite{Dieny91,LTag99}and for technical convenience, we add
to the left external layer of the system one extra layer of a
magnetic insulator (MI), whose role is to pin the direction of the
magnetization $\bf{M}$ in the outer F layer. One of the possible
consequences of that is the return of our system to the initial
state (in a magnetic sense) after switching off the magnetic
field. Otherwise it is necessary to use additional sequences of
the switching-off fields\footnotemark[44] to achieve that. Note
that it is more convenient to use the 4-layered system than the
F/S superlattice~\cite{OurPRB01,OurJETPL01,OurUFN02} since it is
easier to change the mutual ordering of the magnetizations of the
F layers for this system. Thus, formally, our system becomes the
MI/F/S/F$'$/S$'$ one. However this practically  does not affect
the preceding computations performed for the four-layered system,
and we will use the notations earlier introduced for the
four-layered system.

We can control the state of the F/S structure by applying a small
external magnetic field $\mathbf{H}$, which slightly changes the
phase diagram of the sample at the fixed
temperature~\cite{Buz_Ved,LTag99,OurPRB01,OurJETPL01}. In this
case there are specific values of the magnetic field: the
coercivity $H_{\text{coer}}$ at which the orientation of the
magnetization $\bf{M}'$ in the F$'$ layer can be reversed, the
critical field $H_{\text{c}}$ which destroys superconductivity,
and the pinning field $H_{\text{p}}$ at which the direction of the
pinned magnetization $\bf{M}$ in the outer F layer can be
reoriented. Assume that the ordering of localized spins in the F
and F$'$ layers is of an uniaxial type and the magnitudes of these
fields are related as follows~\footnote{By
estimates\cite{Kittel,LTag99,Rusanov_2004PhRvL} $H _
{\text{coer}}\simeq 10\div 100$~Oe. In the case of ferromagnets
with the easy plane parallel to the F/S interface plane $H _
{\text{coer}}\simeq 0$~Oe.
%Note, the phase
%diagrams of the F/S/F$'$/S$'$ structure are practically unchanged
%under action of weak magnetic field $H\simeq$500~Oe. The effect of
%this field $H$ on spin freedoms of conduction electrons is
%negligible in comparison with the exchange field $I\simeq$15000~Oe
%for Gd. The orbital pair-breaking of such external field $H$ also
%extremely small in comparison with upper critical field
%$H_{\text{c2}}2 \simeq$ 1000~Oe for Nb.
}: $H_{\text{coer}} < H_{\text{c}} < H_{\text{p}} $.

The study of the combined phase diagram
(Fig.~\ref{fig5}) helps in optimizing the choice of
parameters of the four-layered F/S/F$'$/S$'$ system,
making it possible to control its superconducting
and magnetic states. Assume the system is in one of
the ``working'' points shown in the diagram. Each of
these ``working'' points characterizes a sample that
is described by a concrete set of parameters
($d_{\text{f}}$, $\sigma _ {\text{s}} $, $2I \tau
_{\text{f}}$, etc.), including the current
temperature of a sample $T$. At zero magnetic field
the system is in the initial \textsf{AFM} state. By
applying an external magnetic field, we can change
the state of the system. The changes in the magnetic
field lead to transitions of the system between
these states. Note that the system in the shown
points ($\star$, $\diamond$, $\vartriangle$) can
have a number of logically different states (up to 7
ones in the $\vartriangle$ case).

Choose one of them, the ``star'', which is located below the
curves $b'$ and $d$  in Fig.~\ref{fig5} (the reduced ``working''
temperature of the sample $t_\star = T_\star/T_{\text{cs}} \approx
0.24$, $\widetilde{d}_\star \approx 0.9$). The system in this
point can have up to 6 different states. Changing first the
external magnetic field $H$ applied in the direction of the
pinning field, one can induce transitions of the system between
the ground \textsf{AFMS} state, the mixed \textsf{FM} state, and
the normal \textsf{FM} state. Applying the external magnetic field
in the opposite direction one can induce three other transitions
between the ground \textsf{AFMS} state, the mixed \textsf{AFM}
state, the antiferromagnetic normal (\textsf{AFMN}) state, and,
finally, the ferromagnetic normal (\textsf{FMN}) one.

\begin{figure} [htbp]
 \centerline {\includegraphics [scale=.45]{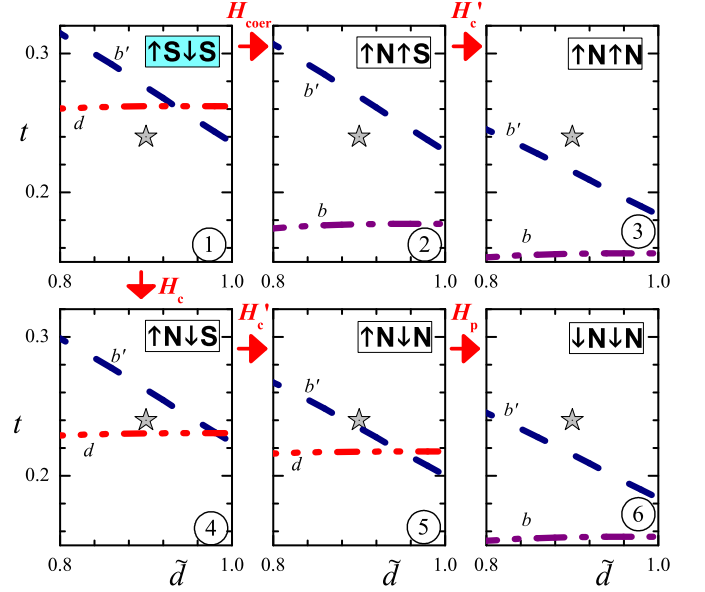}}
%\centerline {\includegraphics [scale=.45]{Figa_Star_F.eps}}
 %\centerline {\includegraphics [scale=.4]{FigaNew.eps}}
 \caption{(Color online)
The qualitative scheme of the chain of phase transitions in the
system in the initial "star" state under the influence of the
external magnetic field $\bf{H}$. Only the envelope phase curves
corresponding to the winning states are shown in each panel for
the S and S$'$ layers. Panel 1 corresponds to the ``star'' point
vicinity in Fig.~\ref{fig5} with $H=0$. Panels 2,3 (4-6)
correspond to the ``upward" (``downward") orientation of $\bf{H}$.
All parameters and notations correspond to the ones used in
Figs.~\ref{fig5} and \ref{fig4}. } \label{fig_Star}
\end{figure}

To show this in more detail, we assume that the
orientation of the magnetization of the outer F
layer pinned, for example, upwards ($ \uparrow $) as
shown in Fig.~\ref{fig1}. At $H=0$ the system is in
the initial \textsf{AFMS} state (see panel~1 of
Fig.~\ref{fig_Star}). If we apply the small external
magnetic field $\mathbf{H}$ that is larger than
coercivity ($H_{\text{coer}}<H<H_{\text{c}}$) in the
direction of the magnetization $\mathbf{M}$ of the F
layer and the pinning field ($\mathbf{H} \uparrow
\uparrow \mathbf{H}_{\text{p}}$), then the direction
of the magnetization $\mathbf{M}'$ in the F$'$ layer
is turned up. The system is transferred into the
state with the ferromagnetic ordering of the
magnetizations, and the \textsf{AFM} state curves
($c$ and $d$) disappear from the diagram. The
remaining curves ($a=a'$, $b'$, and $b$) are only
slightly changed (panel~2 of Fig.~\ref{fig_Star}).
As a result, the data stored using the
superconducting property of the S$'$ layer is kept
unchanged while the information stored on the basis
of the orientation of the magnetizations in the F
and F$'$ layers and the supercurrent in the middle S
layer is changed.

If the applied field $H$ is greater than the
critical field for the outer S$'$ layer
($H>H_{\text{c}}'$), the superconductivity is
destroyed, and the system undergoes a transition
into the \textsf{FMN} state (panel~3 of
Fig.~\ref{fig_Star}). In other words, the
information written using the S$'$ supercurrent
changes as well. Note that one can call the
transition from superconducting to normal state that
is controlled by the external magnetic field to be
the transition with practically infinite
magnetoresistance~\footnote{In this case for the
transition between superconducting ($H=0$) and
normal ($H>H_{\text{c}}'$) states we formally have
$(\rho(H)-\rho(0))/\rho(0)=\infty$.}.

Thus, these transitions can be written as follows
(see also panels 1-3 of Fig.~\ref{fig_Star} for
details): $\uparrow $\textsf{S}$ \downarrow
$\textsf{S} $\xrightarrow{H_{\text{coer}}}$ $
\uparrow $\textsf{N}$ \uparrow $\textsf{S}
$\stackrel{H_{\text{c}}'}{\longrightarrow}$ $
\uparrow $\textsf{N}$ \uparrow $\textsf{N}.

Applying the external magnetic field in the opposite
direction ($\mathbf{H} \downarrow \uparrow
\mathbf{H}_{\text{p}}$) it is possible to induce the
other three additional transitions of the system. In
principle it is necessary to distinguish the
critical fields for the S and S$'$ layers. Moreover
these fields can be different for the same
four-layered samples in the \textsf{AFM} and
\textsf{FM} configurations. For the ``star'' working
point and the \textsf{AFM} ordering, we have to put
that $H_{\text{c}}'> H_{\text{c}}$, since the
difference $[t'_{\text{c}}(\pi 0)-t_{\star}]$ is
larger than $[t_{\text{c}}(\widetilde{\pi \pi}) -
t_{\star}]$  (in Fig.~\ref{fig5} and in panel~1 of
Fig.~\ref{fig_Star} the $b'$ curve is above the $d$
one at $\widetilde{d}=\widetilde{d}_\star$). Note
that generally the required magnitudes of the
corresponding critical fields $H _ {\text{c}} $ and
$H' _ {\text{c}} $ are determined by an appropriate
choice of the working point position relative to the
curves of the superconducting transition $T _
{\text{c}} (d _{\text{f}}) $ and $T' _ {\text{c}} (d
_{\text{f}}) $, respectively (see Fig.~\ref{fig5}).
In addition, by changing $T$ and $d_{\text{f}}$ one
can be always made $H _ {\text{c}} $ smaller than
the field $H _ {\text{p}}$, which is necessary to
remove pinning.

If $H$ is a bit larger than $H_{\text{c}}$ but is
less than $H_{\text{c}}'$ the system undergoes a
transition from the ground \textsf{AFMS} state into
the mixed {\sf AFM} state: $ \uparrow $\textsf{S}$
\downarrow $\textsf{S}
$\stackrel{H_{\text{c}}}{\longrightarrow}$ $
\uparrow $\textsf{N}$ \downarrow $\textsf{S} (see
also panels 1,4 of Fig.~\ref{fig_Star}). Only the
information that is stored using the superconducting
property of the S layer changes, while the stored
data associated with ordering of the magnetizations
of both F and F$'$ layers and the S$'$ supercurrent
remain unchanged. The final transitions in this
series take place at further increase of the
magnetic field (from being $H_{\text{c}}'<H < H _
{\text{p}} $ to $H
> H _ {\text{p}} $) in the same ``downward'' direction: $\uparrow
$\textsf{N}$ \downarrow $\textsf{S}
$\stackrel{H_{\text{c}}'}{\longrightarrow}$ $\uparrow $\textsf{N}$
\downarrow $\textsf{N} $\stackrel{H_{\text{p}}}{\longrightarrow}$
$ \downarrow $\textsf{N}$\downarrow $\textsf{N} (see also panels
4-6 of Fig.~\ref{fig_Star}).

Thus the F/S/F$'$/S$'$ system prepared in such a way has
\textbf{\textit{six}} logically different states: $ \uparrow
$\textsf{S}$ \downarrow $\textsf{S}, $ \uparrow $\textsf{N}$
\uparrow $\textsf{S}, $ \uparrow $\textsf{N}$ \uparrow
$\textsf{N}, $\uparrow $\textsf{N}$ \downarrow $\textsf{S}, $
\uparrow $\textsf{N}$ \downarrow $\textsf{N}, $ \downarrow
$\textsf{N}$ \downarrow $\textsf{N}  (see also
Fig.~\ref{fig_Star}). Recall that in the case of the symmetric F/S
superlattice the superconducting properties vary
\emph{synchronously} in all S layers, and a similar device on its
base has up to \textbf{\textit{five}} logically different
states~\cite{OurPRB01,OurJETPL01,OurUFN02,OurSUST02}. Note also
that only \textbf{\textit{two}} states were proposed for the
trilayer spin switch~\cite{Buz_Ved,LTag99}.

Consider another choice of parameters ($t_\diamond
\approx 0.24$, $\widetilde{d}_\diamond \approx 1.4
$) in Fig.~\ref{fig5} that corresponds to the
``diamond'' working point with four operating
states. At zero magnetic field the system is in the
region that corresponds to the mixed \textsf{AFM}
state (see Fig.~\ref{fig4}), for which the inner S
layer is superconducting, and the outer S$'$ one is
normal. In this case a ``short'' chain of
transitions can be obtained by changing the directed
``upwards'' magnetic field (i.e. $\mathbf{H}
\upuparrows \mathbf{H}_{\text{p}}$):  $ \uparrow
$\textsf{S}$ \downarrow $\textsf{N}
$\xrightarrow{H_{\text{coer}}}$ $ \uparrow
$\textsf{N}$ \uparrow $\textsf{N}. The ``longer''
chain of transitions is obtained when the direction
of magnetic field is changed to the opposite (i.e.
$\mathbf{H} \downarrow \uparrow
\mathbf{H}_{\text{p}}$):  $ \uparrow $\textsf{S}$
\downarrow $\textsf{N}
$\stackrel{H_{\text{c}}}{\longrightarrow}$ $
\uparrow $\textsf{N}$ \downarrow $\textsf{N}
$\stackrel{H_{\text{p}}}{\longrightarrow}$ $
\downarrow $\textsf{N}$ \downarrow $\textsf{N}.
Thus, there are altogether only
\textbf{\textit{four}} different states.

Finally, if we choose the ``triangle'' working point
($t_\vartriangle \approx 0.17$, $\widetilde{d}_\vartriangle
\approx 0.43 $) in Fig.~\ref{fig5}, it is possible to obtain up to
\textbf{\textit{seven}} logically different states. The successive
change of the magnitude of the directed ``upwards'' magnetic field
leads to the following chain of transitions of the system: $
\uparrow $\textsf{S}$ \downarrow $\textsf{S}
$\xrightarrow{H_{\text{coer}}}$
 $ \uparrow $\textsf{S}$ \uparrow $\textsf{S}
 $\stackrel{H_{\text{c}}}{\longrightarrow}$
 $ \uparrow$\textsf{N}$ \uparrow $\textsf{S}
$\stackrel{H_{\text{c}}'}{\longrightarrow}$
 $ \uparrow$\textsf{N}$ \uparrow $\textsf{N}.
 Changing the direction of the
field to the opposite one can obtain the following chain:
$\uparrow $\textsf{S}$ \downarrow $\textsf{S}
 $\stackrel{H'_{\text{c}}}{\longrightarrow}$
  $ \uparrow$\textsf{S}$\downarrow $\textsf{N}
$\stackrel{H_{\text{c}}}{\longrightarrow}$
 $ \uparrow$\textsf{N}$\downarrow $\textsf{N}
$\stackrel{H_{\text{p}}}{\longrightarrow}$
 $ \downarrow$\textsf{N}$ \downarrow $\textsf{N}.
 As it follows from the phase diagram in Fig.~\ref{fig5},
we have used the assumption that $H_{\text{c}} <
H'_{\text{c}}$ for the \textsf{FM} configuration and
$H_{\text{c}}
> H'_{\text{c}}$ for the \textsf{AFM} one
while considering the last working point.
Moreover, if the ``triangle'' position were moved a little
to the left ($\widetilde{d}_{\vartriangle ,{\text{new}}}
\approx 0.42$), then $H_{\text{c}}\approx H_{\text{c}}'$
  for the FM configuration
and we would get the chain of transitions consisting of six
different states.

It follows from our studies that the ``spin switch" device
proposed on the basis of a F/S/F
trilayer~\cite{Buz_Ved,LTag99} (even with the optimum set
of parameters that we have found) has much less number of
logically different ways of data recording than the studied
above four-layered F/S/F$'$/S$'$ system.

\section{Conclusions}

The four-layered F/S/F$'$/S$'$ system has been consistently
studied within the modern theory of the proximity effect with a
detailed account of the given boundary conditions. Theoretical
studies of the critical temperature dependence on the thicknesses
of the F layers have been performed for a wide range of
parameters, and a physically interesting range of their values has
been determined. The latter should be of help in choosing
materials and technology for preparation of the F/S systems with
predetermined properties.

It has been shown that when the $\pi$ phase superconductivity
coexists with the nonequivalence of all layers the physics of the
four-layered systems is considerably richer in comparison with one
for the the earlier studied three-layered F/S/F
system~\cite{Buz_Ved,LTag99} and even the F/S
superlattices~\cite{OurJETPL01,OurPRB01,OurUFN02}. The extra $\pi$
phase superconducting states obtained for the four-layered
F/S/F$'$/S$'$ system have been found to be different from the
analogous $\pi$ phase superconducting superlattice states. The
hierarchy of critical temperatures has been shown to manifest
itself mainly through the occurrence of the different critical
temperatures in the different S and S$'$ layers (space-separated
or decoupled superconductivity). This prediction can be
experimentally verified both for the common case and for the
specially prepared field-cooled \textsf{FM} and \textsf{AFM}
samples.

Theoretical studies performed in this paper have shown that the
four-layered F/S/F$'$/S$'$ system has the best prospects for its
use in superconducting spin electronics (\textit{superconducting
spintronics}). This system can be used for a creation of the
nanoelectronics devices combining within the same layered sample
the advantages of the superconducting and magnetic channels of
data recording that are associated with the conducting properties
of both S and S$'$ layers and the magnetic ordering of the
magnetizations of the ferromagnetic layers. It has been emphasized
that both these channels can be \textit{separately} controlled by
a external magnetic field. These magnetic fields can be made
sufficiently weak~\footnotemark[46] due to a choice of materials
and parameters of the system.
% for essential change of the phase diagrams of
A few versions of the principal scheme of such a
four-layered F/S/F$'$/S$'$ device have been proposed
and explored. It has been shown that the proposed
control device can have up to
\textit{\textbf{seven}} different states, and
transitions between these states can be controlled
by a magnetic field. It should be noted that
advantages of such spin devices are also associated
with their small enough sizes (thickness $d _
{\text{f}} $ is about $0.5\div 5$~nm, thickness $d _
{\text{s}} \sim ~25\div 80$~nm), relatively high
switching speed (according to
estimates~\cite{Clinton97} its range is from
$10^{-10}$\,s$^{-1}$ up to $10^{-11}$\,s$^{-1}$
depending on used materials), relatively large
critical currents (they approximately coincide with
currents for isolated
superconductors~\cite{LTag99}). Certainly, low
temperatures at which usual ``cold''
superconductivity is possible would be a condition
for the use of this type of control device. However,
similar superconducting devices on the basis of the
F/S structures with S layers made out of
high-temperature superconducting
materials~\cite{Habermeier_PhSSA_04} should work at
much higher temperatures.%~\cite{}

Finally, it should be noted that certain
simplifications that were introduced in the
performed studies do not impose any serious
restrictions on the applicability of the main
results obtained in the present paper (hierarchy of
critical temperatures, phase diagram in
Fig.~\ref{fig4}, proposed schemes of a control
device, etc.)
%because we pretend only to
that provide qualitative understanding and principal
solutions of the problem. The model approach used in
our studies adequately describes physics of the
F/S/F$'$/S$'$ system. Though taking into account the
multi-mode
approach~\cite{FominovJETPL01,FominovPRB02} and/or
the 3D treatment\cite{OurUFN02,OurUFN03,OurJETPL00}
could change the shape of the curves in the
discussed above diagrams for the \textit{{same
values}} of parameters, it is always possible
\textit{to change} this set of values in such a way
that \textit{{similar graphs}} can be constructed.
The effect of the influence of each parameter change
on the shape of the phase curves was discussed in
the review~\cite{OurUFN02}.

\begin{acknowledgments}
Yu.P. and M.Kh. are grateful to the Max-Planck-Institute for the
Physics of Complex Systems (Dresden) for providing excellent
conditions for fruitful work. Yu.P. thanks Prof. Fulde and the
participants of his seminar for the productive discussion of the
results presented in this paper. This work was supported in part
by RFBR (No 04-02-16761, 05-02-16369), and the University of Texas
at Arlington Research Enhancement Program.
\end{acknowledgments}

\newpage
%GATHER{proshin.bib}
\bibliography{proshin05}
%\nocite{*}

\end{document}